\documentclass[acmsmall,screen,nonacm]{acmart}

\usepackage{xspace}
\usepackage{soul}
\usepackage{caption}
\usepackage{subcaption}
\usepackage{graphicx}
\usepackage{ifthen}
\usepackage[many]{tcolorbox}
\usepackage{enumitem}
\usepackage{wrapfig}
\usepackage{moresize}
\usepackage{fontawesome}
\usepackage{xstring}

\newboolean{showcomments}
\setboolean{showcomments}{true}

\ifthenelse{\boolean{showcomments}}
{\newcommand{\nbc}[3]{
 {\colorbox{#3}{\bfseries\sffamily\scriptsize\textcolor{white}{#1}}}
 {\textcolor{#3}{\sf\small$\blacktriangleright$\textit{#2}$\blacktriangleleft$}}
 }
}
{\newcommand{\nbc}[3]{}
 }

\newcommand\numchallenges{five\xspace}
\newcommand\tinyskip{\vspace{1.5pt}}

\newtcolorbox{mybox}[2][]{
top=0.15in,left=4pt,right=4pt,bottom=4pt,
fonttitle=\bfseries,
colbacktitle=gray,
colback=gray!5,
colframe=gray!40!black,
enhanced,
attach boxed title to top left={xshift=1.5em,yshift=-\tcboxedtitleheight/2},
boxed title style={size=small},
drop shadow={black!50!white},
title=#2,#1}

\newtcolorbox{challengebox}[2][]{
top=0.15in,left=4pt,right=4pt,bottom=4pt,
fonttitle=\bfseries,
colbacktitle=gray,
colback=white,
colframe=gray!40!black,
enhanced,
breakable,
attach boxed title to top left={xshift=1.5em,yshift=-\tcboxedtitleheight/2},
boxed title style={size=small},
drop shadow={black!50!white},
parbox=false, 
title=#2,#1}

\newtcbox{\lifecyclebox}{%
  nobeforeafter, 
  colframe=blue!50!black, 
  colback=blue!5, 
  boxrule=0.5pt, 
  arc=4pt, 
  boxsep=0pt, 
  left=2pt, 
  right=2pt, 
  top=2pt, 
  bottom=2pt,
  height=15pt,
  valign=center
}

\AtBeginDocument{%
  }

\setcopyright{acmlicensed}
\copyrightyear{2018}
\acmYear{2018}
\acmDOI{XXXXXXX.XXXXXXX}

\acmJournal{JACM}
\acmVolume{37}
\acmNumber{4}
\acmArticle{111}
\acmMonth{8}

\begin{document}


\title{Towards AI-Native Software Engineering (SE 3.0): A Vision and a Challenge Roadmap}


\author{Ahmed E. Hassan}
\orcid{0000-0001-7749-5513}
\email{ahmed@cs.queensu.ca}
\affiliation{%
  \institution{Queen's University}
  \country{Canada}
}

\author{Gustavo A. Oliva}
\orcid{0000-0002-5419-9284}
\author{Dayi Lin}
\orcid{0000-0002-4034-6650}
\author{Boyuan Chen}
\orcid{0000-0001-9103-5820}
\email{cse@huawei.com}
\affiliation{%
  \institution{Centre for Software Excellence, Huawei Canada}
  \country{Canada}
}

\author{Zhen Ming (Jack) Jiang}
\orcid{0000-0002-3063-3197}
\email{zmjiang@cse.yorku.ca}
\affiliation{%
  \institution{York University}
  \country{Canada}
}

\renewcommand{\shortauthors}{Hassan et al.}

\begin{abstract}
    The rise of AI-assisted software engineering (SE 2.0), powered by Foundation Models (FMs) and FM-powered coding assistants, has shown promise in improving developer productivity. However, it has also exposed inherent limitations, such as cognitive overload on developers and inefficiencies. We propose a shift towards Software Engineering 3.0 (SE 3.0), an AI-native approach characterized by intent-centric, conversation-oriented development between human developers and AI teammates. SE 3.0 envisions AI systems evolving beyond task-driven copilots into intelligent collaborators, capable of deeply understanding and reasoning about software engineering principles and intents. We outline the key components of the SE 3.0 technology stack, which includes Teammate.next for adaptive and personalized AI partnership, IDE.next for intent-centric conversation-oriented development, Compiler.next for multi-objective code synthesis, and Runtime.next for SLA-aware execution with edge-computing support. Our vision addresses the inefficiencies and cognitive strain of SE 2.0 by fostering a symbiotic relationship between human developers and AI, maximizing their complementary strengths. We also present a roadmap of challenges that must be overcome to realize our vision of SE 3.0. This paper lays the foundation for future discussions on the role of AI in the next era of software engineering.
\end{abstract}

\ccsdesc[500]{Software and its engineering~Software creation and management}

\keywords{AI-native software engineering, Software Engineering 3.0, Intent-driven development, Conversational AI, AI copilots, Large language models (LLMs), Code synthesis, Knowledge-powered models, Human-AI collaboration, SLA-aware runtime}


\maketitle

\section{Introduction}
\label{sec:intro}

\begin{figure}[!htbp]
    \centering
    \includegraphics[width=1.0\linewidth]{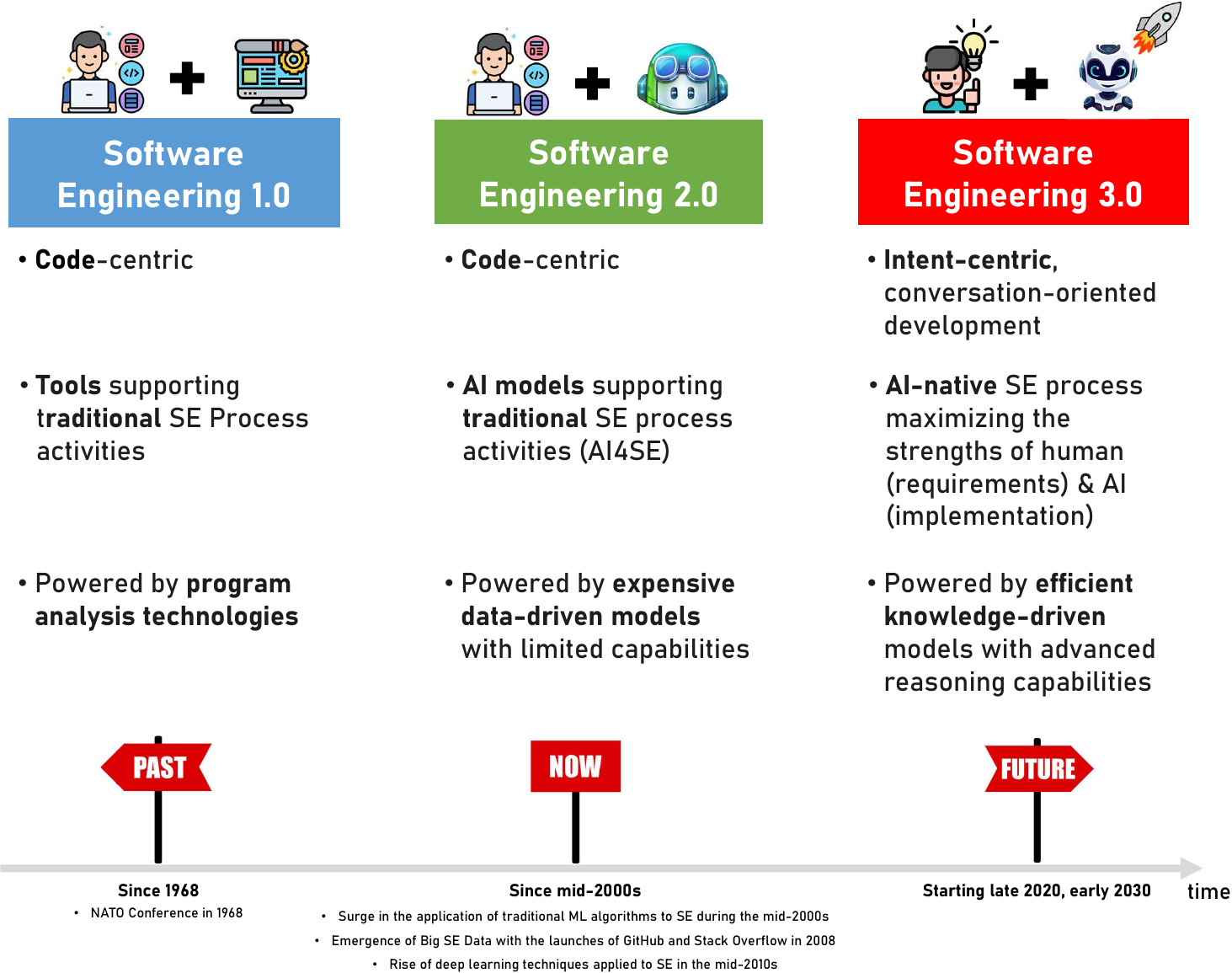}
    \caption{Software engineering evolution.}
    \label{fig:se123}
\end{figure}

We live in the era of AI-assisted SE, which we refer to as \textit{Software Engineering 2.0} (Figure~\ref{fig:se123}). In this era, AI technologies support humans throughout the traditional software engineering process activies (AI4SE). In particular, \textit{AI coding assistants} took the industry by storm. AI coding assistants are tools powered by Foundation Models (FMs), like Large Language Models (LLMs), that help developers write, test, and debug code. AI coding assistants are tightly integrated into developments tools (e.g., IDEs and terminals) and are used by millions of developers around the world on a daily basis. Popular AI coding assistants include GitHub Copilot~\citep{github_copilot}, Claude Code~\citep{claude_code}, Codex CLI~\citep{codex_cli}, Google Gemini Code Assist~\citep{google_gemini_code_assist}, Amazon Q Developer~\citep{amazon_q_developer}, Tabnine~\citep{tabnine}, Cline~\citep{cline}, and Aider~\citep{aider}.

Despite the hype around AI coding assistants, SE 2.0 faces several fundamental limitations. Since humans must still remain at the center of the code-creation loop, developers experience high cognitive overload as they continually decompose problems, prompt AI assistants, evaluate suggestions, debug failures, and iterate through task-driven, code-centric workflows. At the model level, frontier FMs are still trained inefficiently. They depend on massive amounts of unstructured internet data, leading to immense computational and environmental costs, limited depth of reasoning, and ongoing retraining burdens. Finally, AI coding assistants tend to favor additive changes and local optima, often producing bloated or low-quality code that increases complexity, decreases maintainability, and erodes trust in the long term. As AI-generated code becomes pervasive, it also contaminates future training data, creating a feedback loop that further degrades model quality. 

The aforementioned issues call for a deep rethinking of the ways in which we have leveraged AI to engineer software systems. In this paper we introduce our vision of Software Engineering 3.0 (Figure~\ref{fig:se123}). SE 3.0 marks a paradigm shift towards an intent-centric approach, where development is no longer driven by code but by intents (e.g., goals) expressed through iterative, conversation-oriented interactions between human developers and AI teammates. This era leverages the complementary strengths of human developers and sophisticated AI systems. Powered by significantly more efficient AI models with deep SE knowledge, expert-level coding skills, and advanced reasoning capabilities, this era redefines the role of AI in programming and SE. AI evolves from a mere task-driven copilot to an AI teammate that acts as a true partner in a symbiotic relationship with humans. We thus say that SE 3.0 is \textit{AI-native}.

Our SE 3.0 vision and its associated challenges were identified based on (i) surveys of academic and gray literature, (ii) in-depth discussions with industrial and academic leaders (e.g., during \textit{SEMLA 2024}~\citep{SEMLA2024}, \textit{FM+SE Vision 2030}~\citep{FMSEVision2030}, \textit{FM+SE Summit 2024}~\citep{FMSESummit}, the \textit{AIWare} conference series~\citep{AIware} and its associated bootcamps~\citep{AIwareBootcamps}, \textit{Ray Summit 2024}~\citep{RaySummit2024} and the \textit{SE 2030 workshop - FSE 2024}~\citep{SE2030Workshop,SE2030Notes} events), (iii) meetings with our customers and our own internal development teams to understand their frustrations with AI coding assistants and the \textit{status quo}, (iv) our practical experience with the research and development of FMware~\citep{hassan2024rethinking} and the SE 3.0 technology stack (Figure~\ref{fig:se3-stack}), and (v) our close interactions with several industrial partners (40+ leading companies, including Intel, AMD, RedHat, HuggingFace, and SAP) as part of the Open Platform for Enterprise AI (OPEA) alliance~\citep{opeaOpenPlatform}.

This paper is structured as follows. Section~\ref{sec:critical-analysis-se2.0} carries out a critical analysis of SE 2.0. Section~\ref{sec:se3.0} introduces our vision of SE 3.0, focusing on its main principles and supporting technology stack. Section~\ref{sec:challenges} discusses \numchallenges key challenges that must be overcome before SE 3.0 becomes a reality. Lastly, Section~\ref{sec:conclusion} states our conclusions and our final remarks.
\section{A critical analysis of Software Engineering 2.0 (AI-assisted SE)}
\label{sec:critical-analysis-se2.0}

In this section, we introduce Software Engineering 2.0 (Section~\ref{sec:what-is-se2.0}) and discuss its limitations (Section~\ref{sec:what-is-wrong-with-se2.0}).


\subsection{What is Software Engineering 2.0?}
\label{sec:what-is-se2.0}

Software Engineering 1.0 (Figure~\ref{fig:se123}) was predominantly code-centric, with its primary focus on translating requirements directly into code, performing code analysis to ensure quality, and conducting ongoing code maintenance. Several tools were designed to support traditional software engineering activities, such as requirements engineering (e.g., Rational RequisitePro~\citep{RequisitePro2001}, Borland CaliberRM~\citep{BorlandCaliber2006}, and Telelogic Doors~\citep{TelelogicDoors2004,WikipediaRationalDOORS}), design (e.g., Rational Rose~\citep{RationalRose2000,RoseVisualModeling2000,WikipediaIBMRationalRose}, Enterprise Architect~\citep{Bahri2022,WikipediaEnterpriseArchitect}, and ArgoUML~\citep{ArgoUML}), implementation (Eclipse~\citep{EclipseIDE}, Microsoft Visual Studio~\citep{VisualStudio}, and IntelliJ IDEA~\citep{IntellijIDEA}), and testing (e.g., Mercury TestDirector~\citep{TestDirectory2002, MercuryInteractive}, JUnit~\citep{JUnit}, and Selenium~\citep{Selenium}). In the specific context of implementation, several tools leveraged program analysis technologies to help developers manage code complexity (e.g., SciTools Understand~\citep{Understand}), ensure correctness (e.g., FindBugs~\citep{Findbugs}), and optimize system performance (e.g., Mercury LoadRunner~\citep{MercuryLoadRunnerWiki,LoadRunner20024}). Humans were central and drove the entire process. That is, the responsibility for eliciting, interpreting, and realizing requirements, making design decisions, and implementing solutions relied heavily on developers and their ability to use and coordinate between multiple tools.

Software Engineering 2.0 is our current era (Figure 2). It continues the code-centric approach of SE 1.0 while incorporating AI models (e.g., traditional machine learning models~\citep{Lessmann08,Zimmermann09,Tantithamthavorn17}, deep learning models~\citep{Yang22}, and large language models~\citep{Hou24}) to enhance traditional software engineering process activities (AI4SE). Software engineering 2.0 can thus be described as \textit{AI-assisted SE}. AI4SE research largely originated within the Mining Software Repositories (MSR)~\citep{MSR2004,MSRRoadAhead08} community, where researchers first introduced models trained on historical software data and repository-mined patterns, establishing fundamental methods that continue to evolve today. A prime example of AI4SE are AI coding assistants~\citep{github_copilot, claude_code,codex_cli,google_gemini_code_assist,amazon_q_developer,tabnine,cline,aider}, which can write, test, and debug code based on patterns learned from internet-scale datasets and targeted model fine-tuning (e.g., supervised fine tuning, reinforcement learning with verifiable rewards).

\subsection{What are the limitations of Software Engineering 2.0?}
\label{sec:what-is-wrong-with-se2.0}

\subsubsection{High cognitive overload on humans} 

\begin{figure}[!htbp]
    \centering
    \includegraphics[width=0.8\linewidth]{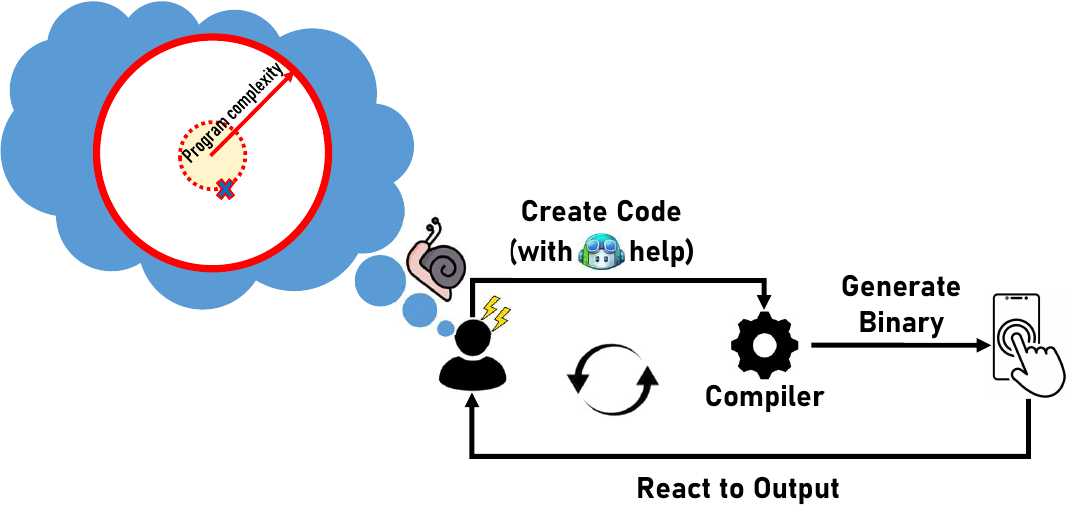}
    \caption{Code-centric development in SE 2.0.}
    \label{fig:se2-dev}
\end{figure}

In SE 2.0, the human developer drives the code creation process loop (Figure~\ref{fig:se2-dev}). As a consequence, the solution space is severely constrained due to the natural limits of the human's cognitive ability to decompose a complex problem, plan possible solutions, and explore those. Also, the final solution must stay within the bounds of human cognitive capability, as developers must be able to maintain and evolve the artifacts (e.g., code).

Humans leverage AI coding assistants to code the many small pieces that make up the final solution. While an experienced developer who uses these assistants will likely be more productive than one who does not, the programming process remains overall inneficient due to the high cognitive overload on the human developer. A typical programming session might look as follows: create a class, write the constructor's signature and have the assistant autocomplete the implementation, create a new method, write a code comment inside the body of that method to induce the AI assistant to generate a piece of code (e.g., ``\texttt{\#\# read the jsonl file}''), rewrite the comment to force the assistant to use some specific package (e.g., ``\texttt{\#\# read the jsonl file using orjson}''), run tests, discover that tests failed because a key requirement was missed, chat with the AI assistant to explain the requirement, reason about the three alternative fixes suggested by the assistant, finish writing the fix, rerun the tests, and so on. As the example highlights, the process is inherently task-driven, code-centric, and puts a significant cognitive overload on the human. In fact, developers have even reported getting trapped in debugging ``rabbit holes'' because of an incorrect or suboptimal solution suggested by the assistant (e.g., the usage of a regular expression for extracting URLs from HTML instead of parsing that HTML and extracting the proper attributes~\citep{Vaithilingam}). Such limitations render the programming process even slower and more cumbersome. 


\subsubsection{Inefficient (and ineffective) model training}
\label{sec:inefficient-model-training}

The training process of frontier FMs, such as OpenAI's GPT 5.2 and Anthropic's Claude Sonnet 4.5, is drastically inefficient due to several factors. These models rely heavily on unsupervised learning, which involves feeding vast amounts of unstructured, internet-scale raw data into the model's architecture. While this approach allows the models to generalize from a wide range of domains and topics, it also leads to significant inefficiencies in terms of computational resources, energy consumption, and data redundancy. The sheer volume of data requires immense processing power, often involving thousands of GPUs running in parallel over extended periods, which drives up both financial and environmental costs. Moreover, the lack of targeted or structured learning means that the models often spend a great deal of time processing irrelevant or noisy information, which contributes little to their final reasoning abilities. In particular, the reliance on raw data limits the ability of these models to effectively capture more specialized or nuanced forms of knowledge that are critical for advanced reasoning. Because the data is largely non-curated, the models must attempt to infer complex relationships, and logic from fragments of information, which leads to incomplete or imprecise understanding in key areas. This often results in models that exhibit surface-level knowledge across a vast number of domains but lack depth or accuracy in more detailed, reasoning-heavy tasks. Additionally, the reuse, evolution, and maintenance of these models over time remain challenging~\citep{Ajibode2025}, as they require continuous retraining and fine-tuning on newer datasets to stay relevant, which compounds the inefficiencies and further increases the complexity of long-term model sustainability.


\subsubsection{Suboptimal code quality and the additive bias in AI coding assistants}
Although early generations of AI coding assistants primarily focused on \textit{adding} code, modern systems are increasingly capable of more diverse transformations. Nevertheless, most assistants still exhibit a strong bias toward additive suggestions rather than prompting developers to pause, rethink, or refactor their design choices. From a software engineering perspective, simply adding more code is not always the most effective solution. Experienced engineers frequently rely on design patterns and refactoring techniques to actually \textit{simplify} or \textit{reduce} code by introducing meaningful abstractions that encapsulate concerns, promote reuse, and improve maintainability. In contrast, an overemphasis on additive changes can contribute to bloated codebases that are harder to maintain, evolve, and modularize. Indeed, a recent study by \citet{HeCMU2025} finds that while Cursor boosts short-term development velocity, it consistently leads to higher code complexity and a sustained accumulation of static analysis warnings, ultimately harming long-term productivity. Consequently, developers often face harder-to-maintain, more convoluted AI-generated code, and increased CI and lint noise. The authors also highlight that multi-file agent edits are still unreliable and erode trust in the tool in the long term. As a result, Cursor's initial speed gains risk being fully canceled out by downstream maintainability problems. A series of studies from Microsoft further indicate that senior developers tend to accept AI-generated suggestions less frequently than junior developers~\citep{ghprodblog, peng2023impact, Ziegler24}. In a survey with almost 500 developers, \citet{Sergeyuk25} investigate the reasons why developers choose not to use AI coding assistants during parts of the development workflow. The top reasons are: lack of need for AI assistance, AI-generated output being inaccurate, developer's lack of trust and the desire to feel in control, and the inability of AI assistants to understand context.


The (i) widespread adoption of AI coding assistants, (ii) the variable quality of assistant-generated code, and (iii) the growing difficulty of distinguishing between AI-produced code and human-written code collectively introduce risks for the quality of the datasets used to train future foundation models. As training data becomes increasingly influenced by assistant-generated artifacts, model outputs may degrade over time, creating a feedback loop in which lower-quality code further contaminates subsequent training cycles. This dynamic threatens to increase the long-term cost of integrating AI into software engineering workflows and underscores the need to transition away from SE 2.0.


\subsection{How about autonomous software engineers?}
\label{sec:devin}


The idea of an autonomous software engineer was popularized by Devin AI~\citep{devinai} and its descendents, including SWE-agent~\citep{SWEAgent}, OpenHands~\citep{OpenHands}, and TRAE~\citep{Trae}. At the time of writing, TRAE ranks first in the \textit{SWE-Bench Verified} leaderboard ~\citep{SWEBenchWebsite}, solving 75.2\% of all benchmark tasks. SWE-Bench verified is a human-validated subset of SWE-bench that more reliably evaluates the ability of an AI to autonomously solve real-world GitHub issues~\citep{SWEBenchVerified,SPICE2025}. 

While TRAE's result is impressive, we believe that the lack of focus on human-AI alignment of intents makes it prone to generate a solution that may not fulfill the original, true requirements. Moreover, current autonomous software engineers still rely on ``off-the-shelf'' FMs for code generation. Despite their ability to generate several candidate solutions to a problem (or part of the problem), the lack of deep SE knowledge and efficiency of the aforementioned models significantly hinder the quality of the code produced by them. Finally, due to the intrinsic limitations of SWE-bench (e.g., only Python projects, only 12 projects, and $\sim$70\% of all tasks originate from 3 out of the 12 projects), the real-world performance of autonomous software engineers remains unclear. Therefore, while we are positive and genuinely curious about the future of autonomous software engineers, we believe that they also come with their own set of challenges that must be overcome before they can be used in a real-world setting. 

As we shall discuss in the next section, the overall philosophy of our vision for SE 3.0 is to leverage the best qualities of humans and AI in a process that is as symbiotic as possible, while also ensuring that a proper technology stack can support such a process.

\section{Our Vision of Software Engineering 3.0 (AI-native SE)}
\label{sec:se3.0}

\begin{figure}[!htbp]
    \centering
    \includegraphics[width=0.8\linewidth]{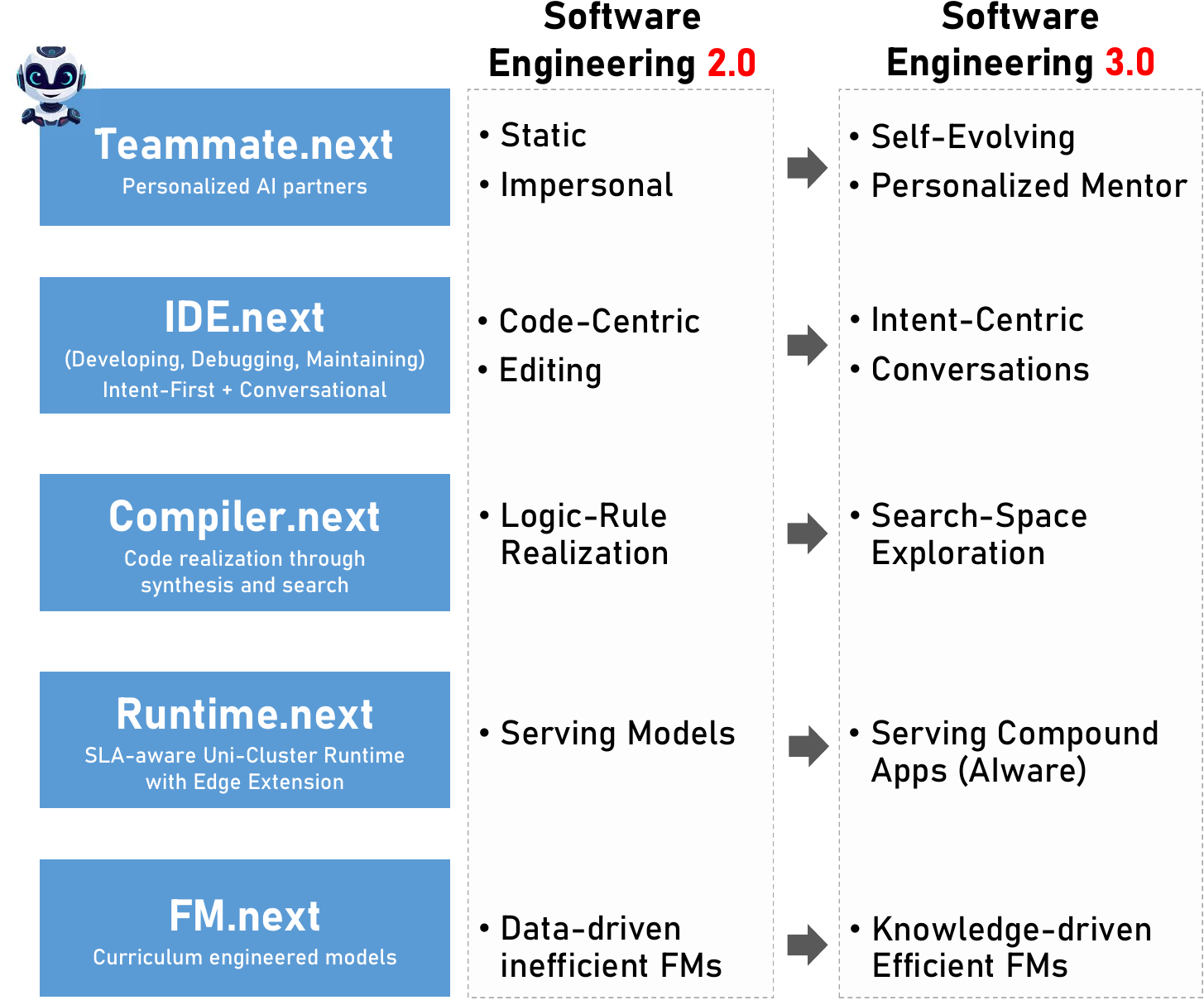}
    \caption{Software engineering 3.0 technology stack (the \textit{.next} suffix serves as an allusion to the future).}
    \label{fig:se3-stack}
\end{figure}

The inherent drawbacks of Software Engineering 2.0 (AI-assisted SE) call for a deep rethinking of the ways in which we have leveraged AI to engineer software systems. In this section, we discuss our vision of a new era of software engineering, which we refer to as Software Engineering 3.0 (AI-native SE). In the next sections, we discuss the core principles of SE 3.0 (Section~\ref{sec:what-is-se3.0}) and the technology stack that supports this new era (Sections~\ref{sec:teammates} to \ref{sec:efficient-models}). Such a technology stack is illustrated in Figure~\ref{fig:se3-stack}.

\subsection{What is Software Engineering 3.0?}
\label{sec:what-is-se3.0}

While SE 2.0 focuses on using AI to support \textbf{traditional} activities (e.g., coding, testing, debugging), Software Engineering 3.0 redefines those activities altogether as well as the technology stack to support them. In SE 3.0, AI is not merely an assistive tool anymore but rather a transformative element (Figure~\ref{fig:se123}). Due to the central and intrinsic role of AI in SE 3.0, we say that SE 3.0 is \textbf{AI-native} (in contrast to AI-assisted). Ultimately, SE 3.0 leverages the complementary strengths of human developers (e.g., ability to reflect about business needs) and sophisticated AI systems (e.g., ability to search for multiple solutions to a problem at \textit{hyper speed}) in the most efficient and effective way possible.
 
SE 3.0 marks a paradigm shift towards an \textbf{intent-centric} approach, where development is no longer driven by code but by \textbf{intents} expressed through back-and-forth \textbf{conversations} between human developers and their AI teammates (conversation-oriented development). We define \textit{intent} as a desired outcome or goal that a human developer wishes to achieve. In SE 3.0, the \textbf{AI drives the code creation loop} by synthesizing the intents into runnable software. Also, as opposed to the expensive data-driven approach for training models (Section~\ref{sec:inefficient-model-training}), SE 3.0 employs a more efficient \textbf{knowledge-driven} approach to training that leverages \textbf{curriculum engineering} to enhance the reasoning capabilities and flexibility of models.

Ultimately, SE 3.0 is a throwback to the \textit{essence} of the SE discipline, which has always been to transform intents into high quality software (e.g., an idea into a mobile app, functional requirements into a system that fulfills a customer's need). \textbf{Code is just a means to an end.}

In the following sections, we discuss the technology stack (Figure~\ref{fig:se3-stack}) that will be necessary to support SE 3.0. Due to the visionary nature of this paper, we focus on its desired attributes instead of its concrete implementation. We expect our vision of this technology stack to serve as a starting point for a broader discussion within the SE community.

\subsection{Teammate.next: From static and impersonal coding assistants to self-evolving personalized mentors} 
\label{sec:teammates}

In SE 3.0, humans collaborate with \textit{AI teammates} instead of AI coding assistants. The AI teammate engages in a conversation with the human developer to help them clarify their intents. Once intents are clarified, the AI teammate synthesizes them into runnable software with the help of \textit{Compiler.next} (Section~\ref{sec:synthesizer.next}). 

However, SE is as much about social interactions as it is about development. As a human partner, the AI teammate exhibits adequate social traits, such as conversational intelligence (e.g., proactivity and communicability), social intelligence (e.g., manners and personalization), and personification (e.g., identity)~\citep{Chaves21}. Such characteristics contrast with AI coding assistants from the SE 2.0 era. For example, if a developer makes the same mistake a thousand times, those assistants will not proactively send a message with skill improvement suggestions or reading recommendations. AI coding assistants are vastly impersonal, almost indifferent.

Another key property of AI teammates is their ability to learn from their mistakes. AI teammates learn not only from human feedback but also \textit{autonomously} from self-reflection and self-exploration. For instance, over time AI teammates learn recurrent contextual information (e.g., project constraints regarding license usage) and human preferences (e.g., regarding coding style) to streamline conversations and speed up the development process. These characteristics again contrast with AI coding assistants, which are mostly static. In turn, teammates can self-evolve through autonomous explorations.

Despite the secondary importance of code in SE 3.0, there will always be codeware, legacy systems to maintain, and low-level code debugging. Hence, AI teammates also serve as \textit{personalized, one-on-one programming mentors}. At any moment, the human developer has the option to ask the AI teammate to explain the rationale and trade-offs behind its coding choices, the design principles and patterns that were used, and even how certain frameworks and libraries work. Indeed, the value of one-on-one mentoring is known to be very high. According to \citet{Bloom84}, an average student performs two standard deviations better with \textit{one-on-one} mentoring as compared to the conventional classroom learning environment. AI teammates can accelerate the growth of junior developers to a senior level more quickly and without any extra cost (e.g., similarly to how Go players became significantly better at the game after integration of AI into the standard gameplay~\citet{Brinkmann2023}).

In summary, Teammate.next represents the adaptive and personalized AI teammate of the SE 3.0 era that will guide and help humans produce runnable software based on their intents. 

\subsection{IDE.next: From code-centric to intent-centric IDEs}
\label{sec:ide.next}

\begin{figure}[!htbp]
    \centering
    \includegraphics[width=0.8\linewidth]{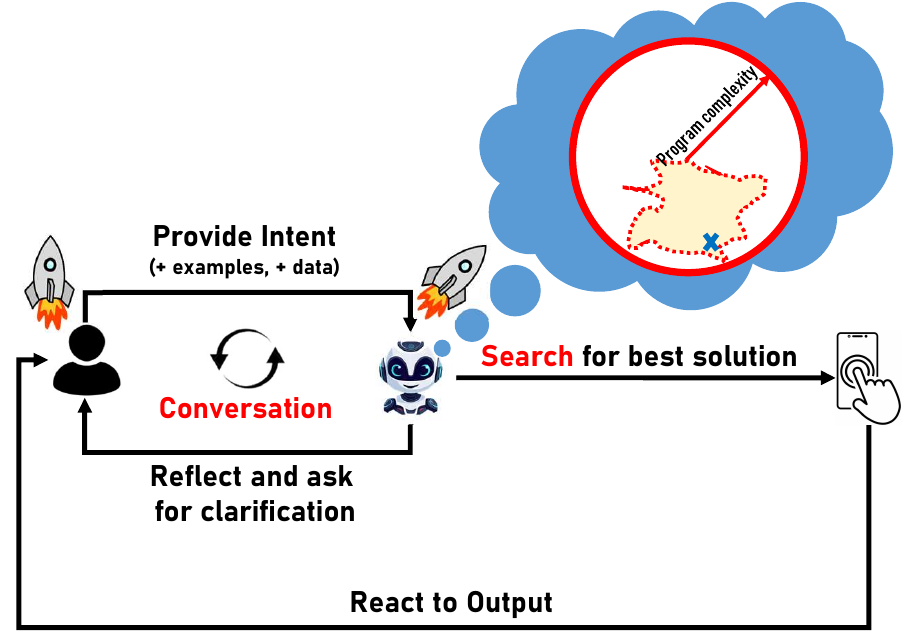}
    \caption{Intent-centric, conversation-oriented development in SE 3.0.}
    \label{fig:se3-dev}
\end{figure}

IDE.next is the new AI-native IDE that powers software development (Figure~\ref{fig:se3-dev}). In SE 3.0, the human developer and his AI teammate first align on \textbf{intents}. In this alignment process, the AI will help the human to both express and refine their intents by means of a back-and-forth conversation. Such a process is needed due to the intrinsic human cognitive inability to fully and clearly express their intents in one-shot. Intents themselves can be described in a flexible manner, ranging from informal descriptions of a functionality to pseudocode examples, UI sketches, and even example data.

Once there is human-AI aligment of intents, the \textbf{AI teammate drives the code creation loop}. As opposed to SE 2.0 (Figure~\ref{fig:se2-dev}), the solution space is now essentially unlimited and can be navigated at hyper-speed by the AI (i.e., coding and human cognition are no longer limitations). Coding thus becomes a search problem, where the fitness function translates to how well the produced code satisfies the human-AI aligned intents. 

The source code itself is of secondary importance and hidden from the human by default. However, the human can always enter a \textit{low-level debugging mode} where they can see the code and adjust it by hand if needed. By means of further back-and-forth conversations between the human and his AI teammate, prototyping loops happen at fast pace. Humans focus on the intents and the AI teammate focuses on turning that into running software. The code synthesis is performed with the help of \textit{Compiler.next}. We discuss the compiler and the AI teammate in more details in Sections~\ref{sec:synthesizer.next} and \ref{sec:teammates} respectively.

Interestingly, the code creation loop can be reinitiated at any time (e.g., when a new FM is released) as long as human-AI conversations are ``archived.'' That is, the \textbf{conversations} become a key asset that needs to be version controlled and managed. We also clarify that we employ the term \textit{code} in its broadest sense possible here. By \textit{code}, we mean traditional code (e.g., Python code), machine learning models, prompts for a FM, and even data (e.g., additional training data to fix incorrect classifications made by a neural network~\citep{TeslaWipers}).

The development model in SE 3.0 is inherently iterative and cyclic, where humans react to the prototyped software in every code cycle. The cycle ends once the human developer is satisfied with the final working software, meaning that all intents have been clarified and adequately fulfilled. This development model draws inspiration from Test-Driven Development (TDD), which (i) compels developers to have a deeper and earlier understanding of the requirements and (ii) emphasizes the role of code as a means, not an end~\citep{beck2000extreme,beck2002test}. 

\subsection{Compiler.next: From logic-rule realization to search-space exploration}
\label{sec:synthesizer.next}

\textit{Compiler.next}~\citep{CompilerNext25} is the piece of software responsible for synthesizing intents (possibly enriched with examples and data as shown in Figure~\ref{fig:se3-dev}) into runnable software via an efficient \textbf{search} process. At hyper speed, the solution is iteratively developed through code mutations and a self-reflection mechanism that evaluates (i) how well the code matches the intent and (ii) the quality of the code (to enable low-level debugging by humans if needed). The synthesizer is powered by efficient and knowledge-driven FMs, which we discuss in Section~\ref{sec:efficient-models}. As such, the synthesizer is an FMware (a software system that uses FMs as one of its building blocks)~\citep{hassan2024rethinking}.

A key attribute of the code synthesizer is its ability to find the best trade-off between multiple competing objectives, such as \textit{accuracy} (how well it fulfills the intent), latency (e.g., number of requests sent to the FM), and cost (e.g., number of input tokens in the prompt and expected number of output tokens). In other words, the synthesizer performs a \textit{multi-objective optimization}. We note that the design and complexities of such a multi-objective optimization are completely hidden from the end-user (developer), who only interfaces with \textit{Teammate.next}.

\begin{figure}[!htbp]
    \centering
    \includegraphics[width=1.0\linewidth]{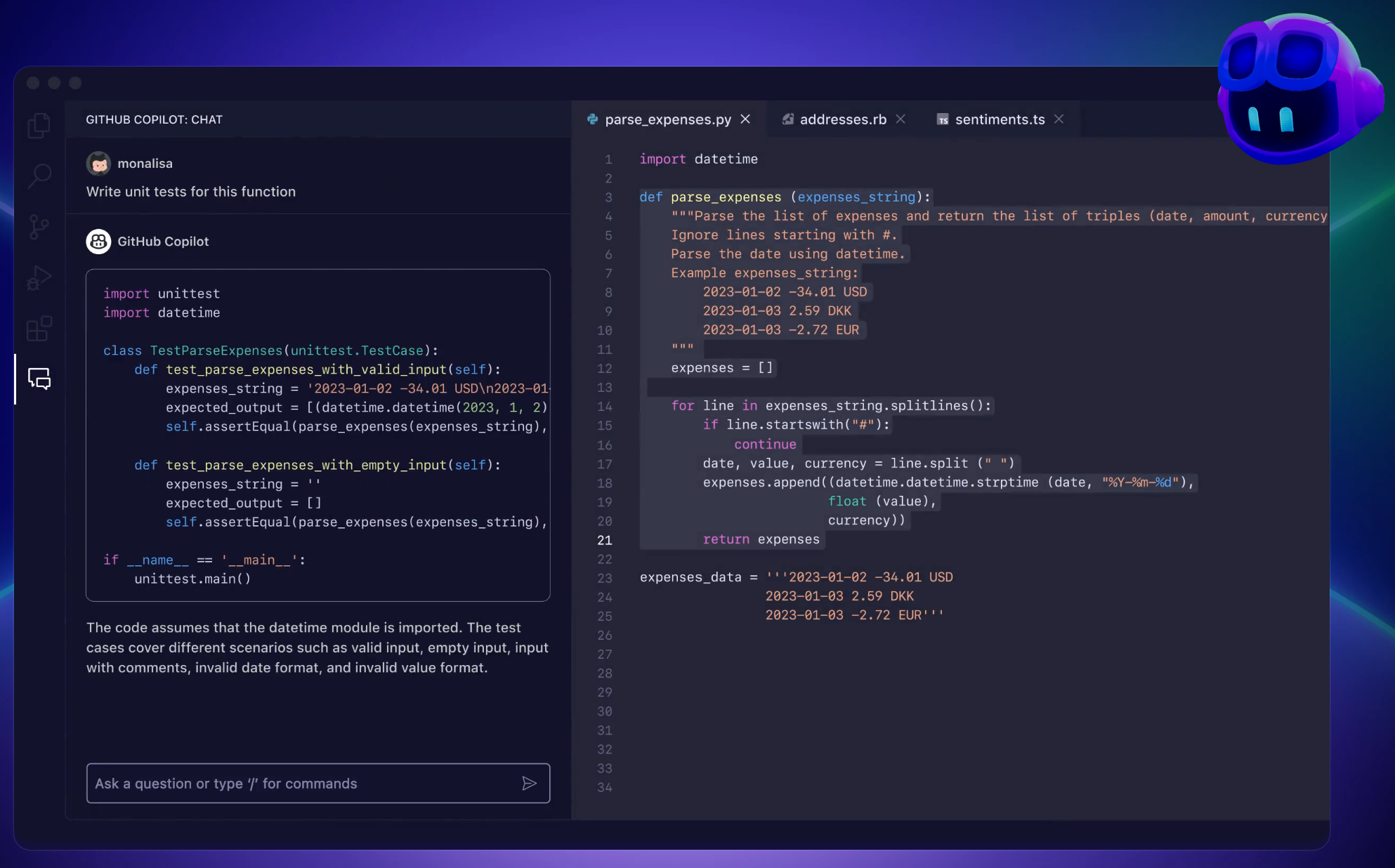}
    \caption{Screenshot from GitHub Copilot's webpage showcasing the copilot's features~\citep{GitHubCopilotWebpage}: ``Write unit tests for this function.''}
    \label{fig:test-from-code}
\end{figure}

Another key attribute of Compiler.next is its \textit{goal-tracking} mechanism. This mechanism ensures that (i) intents are translated into tests, (ii) tests are adequately adapted as requirements change, and (iii) tests will eventually pass. In other words, this mechanism guarantees that tests are always derived from intents (requirements) instead of some existing piece of code. Deriving tests from code is a well-known AI4SE pitfall, since there is no guarantee that the code does what it is supposed to do~\citep{Harman24, Mathews2024}. However, earlier versions of AI assistants still promoted such practice (Figure~\ref{fig:test-from-code}).

In a dedicated paper~\citep{CompilerNext25}, we introduce a conceptual architecture, a proof-of-concept implementation, and an initial empirical evaluation of Compiler.next. At its core, Compiler.next is organized as a modular stack comprising architecture explorers, prompt rewriters, search optimizers, and observability mechanisms, all coordinated by a multi-objective search engine that balances accuracy, cost, and latency. This architecture is realized through a working prototype that implements iterative synthesis, self-reflection, semantic caching, and distributed execution. An initial evaluation on the HumanEval-Plus benchmark~\citep{Evalplus} demonstrates the feasibility of the approach, showing that Compiler.next can automatically synthesize and optimize system prompts from high-level intents while improving accuracy (i.e., solutions that pass all the unit tests associated with each HumanEval-Plus task) and efficiency, thus providing early evidence that intent compilation via search is both practical and effective. In the future, we may implement Compiler.next as a multi-agent system (e.g., using a framework like Microsoft Autogen~\citep{MSAutogen}) to leverage the natural decomposition of complex tasks into modular agent roles. 

\subsection{Runtime.next: From the serving of models to compound apps}
\label{sec:runtime.next}

FMware raised in popularity with the release of not only closed-source FMs (e.g., GPT 5.2) but also open-source ones with better reasoning capabilities (e.g., DeepSeek V3.2). Frontier FMs are also typically multi-modal, meaning they can process not only text but also other forms of input such as images and video, creating several business opportunities for FMware~\citep{Ding24}. In fact, the technology stack that we describe in this paper also heavily relies on FMs (e.g., the \textit{Compiler.next}). Moving forward, several software systems are expected to either be FMware or contain FMware as one of its modules~\citep{hassan2024rethinking}. 

However, it does not stop there. FMware development typically embraces the \textit{data flywheel}~\citep{DataFlywheel} approach. The data flywheel is a self-reinforcing cycle where data collection, analysis, and insights drive continuous improvements and growth in a system. As more data are gathered (e.g., field, telemetry, and human feedback data), it improves the accuracy and effectiveness of algorithms and processes (e.g., model fine-tuning and agent improvement), which in turn enhances user experiences and operational efficiency. This leads to more users or activities that generate even more data, fueling further optimization. As a result, FMware remains in a \textit{continuous state of evolution}.

The aforementioned FMware characteristics put pressure for a new type of runtime. \textit{Runtime.next} represents our vision of a SE-3.0-ready runtime and exhibits the following qualities (Figure~\ref{fig:se3-stack}): 

\begin{itemize}[wide = 0pt, itemsep = 1.5pt, topsep = 1.5pt]
    \item \textbf{SLA-aware.} Similarly to SE 1.0 and 2.0, SLAs will vary drastically depending on the type of FMware. For instance, (i) real-time applications (e.g., chatbots) require minimal latency, especially for the first token generation, to ensure responsiveness during live interactions, (ii) batch processing tasks (e.g., document summarization) typically focus on maximizing throughput while allowing slightly higher latencies, and (iii) memory-intensive tasks (e.g., large input/output processing) require careful memory and resource management to prevent contention and ensure scalability. Towards addressing the SLA requirements of different types of applications, Runtime.next implements intelligent priority-based routing and leverages sophisticated observability machinery to schedule and allocate resources more effectively. Concretely, in a dedicated paper~\citep{Zhang2024}, we introduce the architecture and a prototype implementation of Runtime.next targeting FMware. The runtime models FMware as DAG-based workflows and decomposes the end-to-end SLA into per-task slack, which is a time buffer assigned to each model invocation that can be consumed or saved as execution progresses. A \textit{profiler} estimates task latency and initial slack, a \textit{resource provisioner} elastically scales replicas when slack is at risk, an intelligent \textit{router} dispatches requests based on remaining slack and queue states, and a \textit{cluster manager} enforces these decisions. Empirical evaluations in a real-world deployment scenario showed fewer SLA violations and better hardware utilization than existing non–SLA-aware serving frameworks under increasing load.

    \item \textbf{Uni-clusters.} A uni-cluster unifies multiple FM-related activities — such as training, fine-tuning, serving, and agent self-evolution — within a single cluster~\citep{hassan2024rethinking}. This design drastically simplifies the management of computing resources and lowers operational costs, as it eliminates the need for separate infrastructures for each task. Uni-clusters not only optimize hardware utilization but also streamline operational processes, enhancing the flexibility and productivity of Ops teams. By consolidating various FMware-related activities into one cluster, uni-clusters facilitate smoother scaling and more agile adaptation to changing workloads, further improving the ability to meet SLAs.

    \item \textbf{Edge-computing extension.} To ensure AI teammates are effective for daily use, they must offer low-latency responses and be cost-effective. Large frontier models such as DeepSeek v3.2 have hundreds of billions of parameters and are thus very expensive to deploy and operate. Therefore, a critical feature of \textit{Runtime.next} is its \textit{edge-computing extension}. This extension can redirect simpler requests to a smaller, more efficient model, such as one running locally on the developer's laptop via Ollama~\citep{Ollama,RouteLLM,RouterBench}. The aim of the edge extension is to handle as many requests as possible with the smaller model (reducing both computational costs and latency), while still maintaining a response quality that closely matches that of the larger model. Determining the best algorithm for routing requests remains an open research challenge (Section~\ref{cha:runtime}, OQ4).
\end{itemize}


\subsection{FM.next: From data-driven inefficient FMs to knowledge-driven efficient FMs}
\label{sec:efficient-models}

The lack of a focused and structured learning approach for models in SE 2.0 results in significant inefficiency, as those models often spend a substantial amount of time processing irrelevant or noisy information. This not only consumes valuable resources but also adds little to enhancing their reasoning abilities.

An alternative and more efficient strategy for training FMs is via \textit{curriculums}~\citep{SurveyCurriculumLearning}. \textbf{Curriculum engineering} is the systematic design, development, and continuous refinement of curriculums that contain curated, organized, and high-quality domain-specific knowledge. A curriculum does not need to be manually designed from top to bottom. For instance, humans can leverage AI to kickstart the curriculum then manually revise it. Once a curriculum is stable, humans can again leverage AI to generate synthetic data (e.g., concrete examples) to enrich the curriculum. As an illustrative example, one could enrich the ``design patterns'' section of a SE curriculum by having the AI generate examples of the application of design patterns in different programming languages and in different contexts. In fact, synthetic data plays a pivotal role in overcoming the limitations of real-world data, as it allows for scalable, controlled, targeted, evolvable, and maintainable training datasets. In summary, curriculum engineering values high-quality data curation coupled with synthetic data generation over blind scrapping of internet-scale data (similarly to the \textit{phi} family of models from Microsoft~\citep{PhiModels}).

The FM used by \textit{Compiler.next} is expected to be trained with a high-quality SE curriculum (e.g., by drawing inspiration from the IEEE Software Engineering Body of Knowledge -- SWEBOK~\citep{Washizaki2024}). We refer to it as FM.next. FM.next must show \textit{uniform competence across the entire software engineering spectrum}, such as requirements reasoning, architectural design, implementation, testing, debugging, and maintenance. This holistic competence is essential, since Compiler.next must be able to reason holistically across the entire SE lifecycle within a single intent-to-software loop (e.g., design or refine an architecture, generate tests from clarified goals, debug internal synthesis cycles, make trade-offs).

We also note that a curriculum \textit{externalizes} the knowledge from the model itself, allowing for higher adaptability. For instance, the curriculum can be reviewed, reorganized, or augmented in response to the model's performance on specific tasks. Here, \textit{observability data} becomes essential, particularly \textit{cognitive} observability~\citep{Watson25}. By closely monitoring key performance metrics, such as where the model is excelling or underperforming, it is possible to adaptively adjust the curriculum and synthetic data generation processes to improve FM.next in a systematic manner. This feedback loop ensures that the training process focuses on weaker areas, allowing for a more efficient and focused improvement in model reasoning. This structured approach not only leads to better model performance but also dramatically reduces the inefficiencies that stem from unstructured learning in traditional model training workflows (e.g., the model's proneness to hallucinations). 

Ultimately, curriculum engineering provides a high-level structured vision of how models should be created, maintained, reused, and evolved. In many ways, curriculums represent the \textit{intellectual property} instead of the model itself. Given their fundamental role, curriculums need to be properly asset managed and versioned, such that they can be expanded and reused across models and use cases.

\smallskip \noindent \textbf{A reference recipe for designing and maintaining curriculums.} Curriculum design involves identifying key topics and organizing them in a logical manner, typically in the form of a taxonomy. Researchers have recently started conceiving systematic approaches for curriculum design. A prominent example is IBM's InstructLab~\citep{Sudalairaj2024, InstructLab}. In InstructLab, a curriculum is developed through a collaborative process involving domain experts, data scientists, and AI practitioners. The process begins with defining objectives and scope, identifying key domains and subdomains, and engaging domain experts to outline core concepts, tasks, and expected input-output specifications. The curriculum is then usually structured in a hierarchical fashion, with broad categories at the top and finer-grained subcategories below. Each node represents a specific task or knowledge area with corresponding examples, templates, and evaluation rules. InstructLab suggests branching the curriculum's root node into knowledge, foundational skills, and composition skills (i.e., tasks that require a synergetic combination of knowledge and foundational skills to answer complex queries from users). The limited quantity of manually curated data samples embedded within the leaf nodes of the taxonomy is often inadequate for achieving satisfactory instruction-following performance. To address this challenge, a taxonomy-driven approach to synthetic data generation, such as leveraging a teacher FM, can be utilized. After that, the curriculum is continually tested for internal consistency to ensure that categories and subcategories do not overlap unnecessarily and are mutually exclusive where appropriate. Iterative refinement, pilot testing, and community contributions (e.g., via crowdsourcing) further enhance the curriculum. In particular, a data flywheel approach (Section~\ref{sec:runtime.next}) is recommended for curriculum refinement. As part of this approach, curriculum engineers modify the curriculum (e.g., by adding new topics or examples) to overcome problems identified in observability data. Common sources of problems include underrepresented areas in the curriculum, emerging trends in the data, and concept drift~\citep{ConceptDrift14}). 

\begin{figure}[!htbp]
    \centering
    \includegraphics[width=1.0\linewidth]{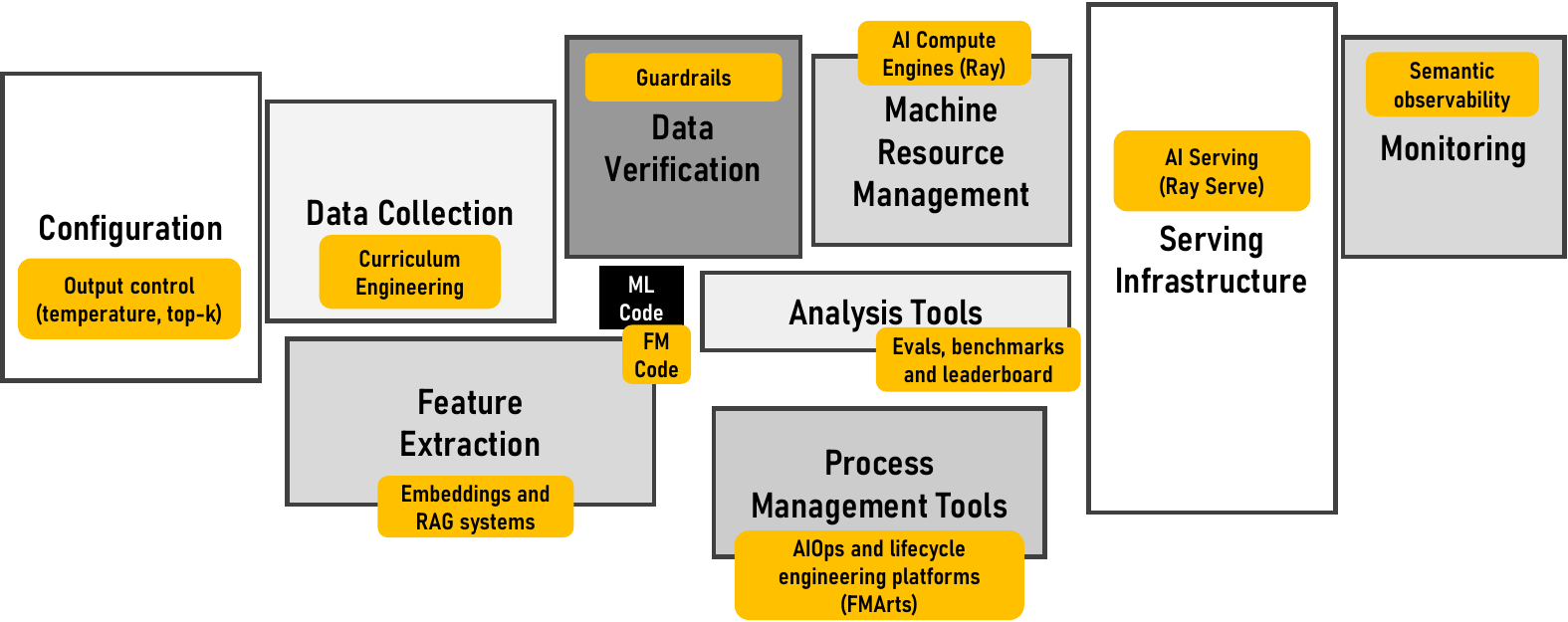}
    \caption{A reinterpretation of the picture by ~\citet{Sculley15} showing the relative size of FM code in the landscape of AI systems enginering in the SE 3.0 era.}
    \label{fig:scully-in-se3-era}
\end{figure}

\smallskip \noindent \textbf{Are knowledge-driven efficient FMs all we need?} We finish this section with a word of \textit{caution}. Despite the unquestionable importance of FMs and the procedures used to train them, the models themselves are only ``one part of the equation'' for enabling efficient and effective FMware development in the SE 3.0 era (Figure~\ref{fig:scully-in-se3-era}). Overfocusing on the model itself is a repetition of a well-known AI engineering pitfall from the SE 2.0 era~\citep{Sculley15}.

\section{Challenges}
\label{sec:challenges}


In this section, we introduce key challenges in SE 3.0. For each challenge, we include a description, what parts of the SE 3.0 stack it affects (Figure~\ref{fig:se3-stack}), one or more open questions, and our vision regarding the solution to those questions. Overcoming these key challenges requires novel technologies and a deep rethinking of how humans and AI interact, both of which should be driven by cutting-edge software engineering research. The list of challenges that we present is not meant to be extensive, but rather focuses on the major painpoints that we observe based on our practical experience and discussion with academic and industry leaders.  

\subsection{Speeding up human-AI alignment}
\label{cha:human-ai-alignment}

\noindent \textbf{Description.} Humans have a limited ability to fully and clearly express their intents in one-shot using written text (e.g., humans may forget to communicate important aspects of a requirement). As a consequence, human-AI alignment of intents becomes slow. We note that this is a problem that will continue to exist no matter how powerful FMs may become, since it stems from an intrinsic human limitation. That is, humans are the bottleneck, not the AI. To add to the challenge, natural language is inherently ambiguous~\citep{JurafskyMartin2009Chapter19}.

\tinyskip \noindent \textbf{Affects.} IDE.next, Teammate.next

\begin{challengebox}{Open question \#1}
    \textbf{OQ1:} How to balance between asking \textit{too many clarifying questions} and \textit{not asking enough}?
    
    \medskip \noindent \textbf{Our vision.} AI teammates should have the ability to develop a \textit{theory of the mind} (ToM) of the human with whom they are interacting. ToM is a psychological concept that refers to one's ability to understand and attribute mental states to others~\citep{BaronCohen85}. At its core, ToM is about recognizing that other people have their own thoughts and perspectives that may be different from our own. Such a recognition is crucial for social interactions, as it helps us predict and interpret the behavior of others. ToM is developed in early childhood and plays a critical role in our ability to communicate and form relationships.

    We draw the following analogy: the manner in which a task is described to a new development team member is typically more comprehensive compared to when that task is assigned to a senior team member. The senior team member already has assumptions in their head regarding the expectations of the task assigner (especially when they have had a history of interactions). Similarly, an AI teammate should learn from prior interactions with humans, such that intent details can be omitted, recurrent themes do not require repeated clarifications, and missing information can be inferred. Therefore, ToM must be ``sticky'' over interactions yet adjustable to the specific context.
    
    Designing effective mechanisms for enabling an AI teammate to develop ToM remains an open challenge. At the very least, such a mechanism would need to store prior conversations with the human, continuously summarize them, and inject them as appropriate into the context window of the underlying FM. The teammate is also expected to leverage linguistic clues from developers' questions (e.g. readability, emotion, richness of wording, adaptability to AI answers) to understand the developer's perception of itself (e.g., anthropomorphic, intelligent, favorable)~\citep{ToM2}. However, even with a perfect theory of someone's mind, there will likely be important contextual factors that the human will forget to mention or even consider (again, an intrinsic human limitation). Hence, the teammate should constantly strive to refine the human's intents.
    
    Interestingly, humans also require \textit{signals} from the AI teammate (e.g., textual feedback, confidence scores) to be able to judge whether it actually understood the communicated intent. This concept, recently termed \textit{mutual ToM}~\citep{wang2024mutualtheorymindhumanai}, underscores the necessity for humans to also develop a theory of the ``mind'' of AI. Such signals are needed due to natural language ambiguity as well as inherent limitations of a text-based interaction model~\citep{interactiondesign} (e.g., lack of visual cues such as facial expressions). Current AI assistants do not indicate by default how confident they are in their answers.

    An empirical study conducted by our group~\citep{Gallaba25} provides complementary evidence for this vision: a multi-agent, ToM–enhanced conversational system significantly improved intent clarification, requirement completeness, and human–AI alignment across 150 diverse scenarios, demonstrating that ToM-oriented reasoning and iterative refinement can clearly strengthen the AI teammate’s ability to understand and operationalize human goals.

    Finally, we clarify that ToM \textit{is not a silver bullet} and that other complementary techniques and approaches are likely required to speed up intent-alignment and the overall conversation efficiency. In particular, requirement engineering activities remain key in SE 3.0 (e.g., customer feedback on software product versions should inform conversation-oriented development).
    
\end{challengebox}

    
    

\subsection{Improving the efficiency of code synthesis}
\label{cha:solution-synthesizer}

\noindent \textbf{Description.} Code synthesis is a key component of intent-centric, conversation-oriented development in SE 3.0 (Figure~\ref{fig:se3-dev}), which will be executed many times by developers on a daily basis. As such, synthesis needs to be as efficient as possible.

\tinyskip \noindent \textbf{Affects.} Compiler.next, Teammate.next

\begin{challengebox}{Open question \#2}
    \textbf{OQ2:} How can we improve the efficiency of the code synthesis process while still preserving (or ideally even increasing) its accuracy?

    \medskip \noindent \textbf{Our vision.} A promising approach is to enhance the efficiency of the code synthesis process by leveraging principles from Search-Based Software Engineering (SBSE)~\citep{Harman2012}. Instead of initiating every search from scratch, we propose using past search data (both from local and crowdsourced runs) to guide future search processes. By caching previous search outcomes, the synthesizer can reuse and build on prior knowledge, reducing redundant computations and allowing it to start searches closer to optimal solutions. This reuse of search results not only saves time but also allows for smarter exploration of the problem space. Additionally, extrapolating insights from historical data can help tune search algorithms, such as genetic algorithms or simulated annealing, making them more efficient for specific code synthesis tasks.

    Beyond simply reusing results, historical search data can inform the creation of new heuristics tailored to particular problem domains. By analyzing patterns and characteristics of past successful searches via \textit{self-reflection}, the synthesizer can learn new heuristics that improve the search process, making it more targeted and accurate. This would allow the system to better adapt to different kinds of code synthesis problems and provide more efficient solutions that are aligned with user needs.
    
    Finally, incorporating past search data allows for a degree of personalization in the code synthesis process. By learning from a user's previous search behaviors and preferences, the searches can be guided toward more relevant solutions, increasing both the efficiency and quality of the results. Over time, as more data is accumulated, the system will continuously improve, offering increasingly efficient and accurate code synthesis.

    This vision is further supported by our complementary work on Compiler.next, which demonstrates through a search-based compilation framework and empirical validation that reusing compilation traces, leveraging semantic caching, and integrating self-reflective search heuristics can substantially improve both the efficiency and accuracy of FM-powered code synthesis systems~\citep{CompilerNext25}.
\end{challengebox}


\subsection{Improving runtime performance}
\label{cha:runtime}

\textbf{Description.} Runtime is a critical piece of the SE 3.0 stack. As FMware gets more and more complex, efficiently serving and evolving these applications while meeting SLAs and making the best use of heterogeneous hardware resources remains a key challenge. 

\tinyskip \noindent \textbf{Affects.} Runtime.next

\begin{challengebox}{Open question \#3}
    \textbf{OQ3:} How can we improve the runtime performance beyond what state-of-the-art serving frameworks such as Ray Serve~\citep{rayserve} can provide?

    \medskip \noindent \textbf{Our vision.} In FMware, the computation sequence is usually described in the form of a workflow that prescribe which models (or agents) will be triggered, how, and which order. While the norm is to use \textit{inductive} expressions (e.g., a Python script) to represent this workflow, our vision is to compile such workflow into a graph representation that allows for a \textit{declarative} expression of the FMware logic and preserves rich \textbf{intent information}. With intents preserved, the runtime becomes able to process this graph to perform several optimizations to decrease FM inference latency, decrease data movement, and maximize resource utilization. 
    
    We have been refining and implementing this vision through FMArts (a lifecycle engineering platform for FMware) and its Fusion Runtime~\citep{hassan2024rethinking}. Preliminary results show a latency improvement in the order of 30\% compared to Ray Serve.
\end{challengebox}


\begin{challengebox}{Open question \#4}
    \textbf{OQ4:} What is the best algorithm for routing requests in the context of the runtime's edge-extension?

    \medskip \noindent \textbf{Our vision.} An analysis of the state-of-the-art~\citep{RouteLLM,RouterBench} aproaches reveals two key limitations. First, they rely on learning user preference from a pre-determined dataset of prompts. However, real-world requests can be out-of-distribution with the profiled data, leading to incorrect routing decisions. In the case of RouteLLM~\citep{RouterBench}, performance depends on the specific type of task. Second, once the routing system is deployed, the learned routing is static (i.e., does not evolve). As a consequence, it may send requests to the more expensive cloud model even if the edge model is capable of handling the request. 

    Inspired by the theories of continual learning~\citep{ContinualLearning} (the process of learning new skills and knowledge on an on-going basis) and self-evolving agents (e.g., Voyager~\citep{wang2023voyager}), we believe that a promising research avenue is to provide useful guidance to the edge model to help it solve unseen tasks via in-context learning. The fundamental idea is that the edge model can learn new skills as it sees more examples. Over time, the edge model should become more capable and handle more requests in a localized manner, leading to a more responsive experience and a reduction of operational costs. A preliminary design, implementation, and evaluation of this idea is discussed in our complementary work (\textit{real-time adapting routing})~\citep{Vasilevski2024}. In particular, on different subsets of the popular MMLU benchmark~\citep{MMLU2021}, our approach routes 50\% fewer requests to computationally expensive models while maintaining around 90\% of the general response quality.

\end{challengebox}
\subsection{Improving FM's understanding of code and SE}
\label{cha:smarter-fms}

\textbf{Description.} Despite substantial advances in frontier FMs for software engineering (e.g., GPT-5.2-Codex), their understanding of software engineering remains incomplete. Frontier FMs can generate and reason about code far more effectively than earlier FMs, yet much of this capability still stems from pattern learning rather than a grounded grasp of program execution, design intent, or real-world development workflows. As a result, they still struggle with certain tasks requiring deeper semantic understanding, long-horizon planning, or integration across the full software engineering lifecycle. This highlights that genuinely robust comprehension of software and its development processes remains an open challenge.

\tinyskip \noindent \textbf{Affects.} Compiler.next, Teammate.next

\begin{challengebox}{Open question \#5}
    \textbf{OQ5:} What approaches should we use to improve FMs so that they not only understand code but also broader software engineering principles?

    \medskip \noindent \textbf{Our vision.} Current FMs increasingly incorporate runtime information, most notably through reinforcement learning with verifiable rewards (e.g., test case results). Yet, we believe that there remain opportunities to explore more nuanced forms of runtime (dynamic) data. These include detailed representations of execution logic~\citep{Ding24}, evolving variable states, and call stacks. Techniques such as symbolic execution and data flow analysis might also help bridge static and runtime data, teaching models to internalize how variables, memory, and control flow evolve during execution. Such an expanded training paradigm should enable FMs to produce code that is not only correct but also more aligned with core software engineering principles and practices.
    
    In addition to multi-modality, we believe curriculum engineering to be a suitable approach to teach models about the broader context in which code is developed, such as system design, version control, testing, debugging, and collaboration across teams. These aspects are critical in real-world software engineering, where the interplay between static code, dynamic behavior, and software development lifecycle processes determines the quality and maintainability of software. Designing an effective SE curriculum remains an open challenge.

    Finally, designing mechanisms to study how LLMs internally represent and understand code is a promising research avenue~\citep{North25}. Such mechanisms can help identify gaps in understanding and guide targeted improvements in model architecture and training data.

\end{challengebox}
\subsection{Eliminating the need for prompt engineering}
\label{cha:from-prompt-to-intent}

\begin{figure}[!htbp]
    \centering
    \fbox{\includegraphics[width=0.8\linewidth]{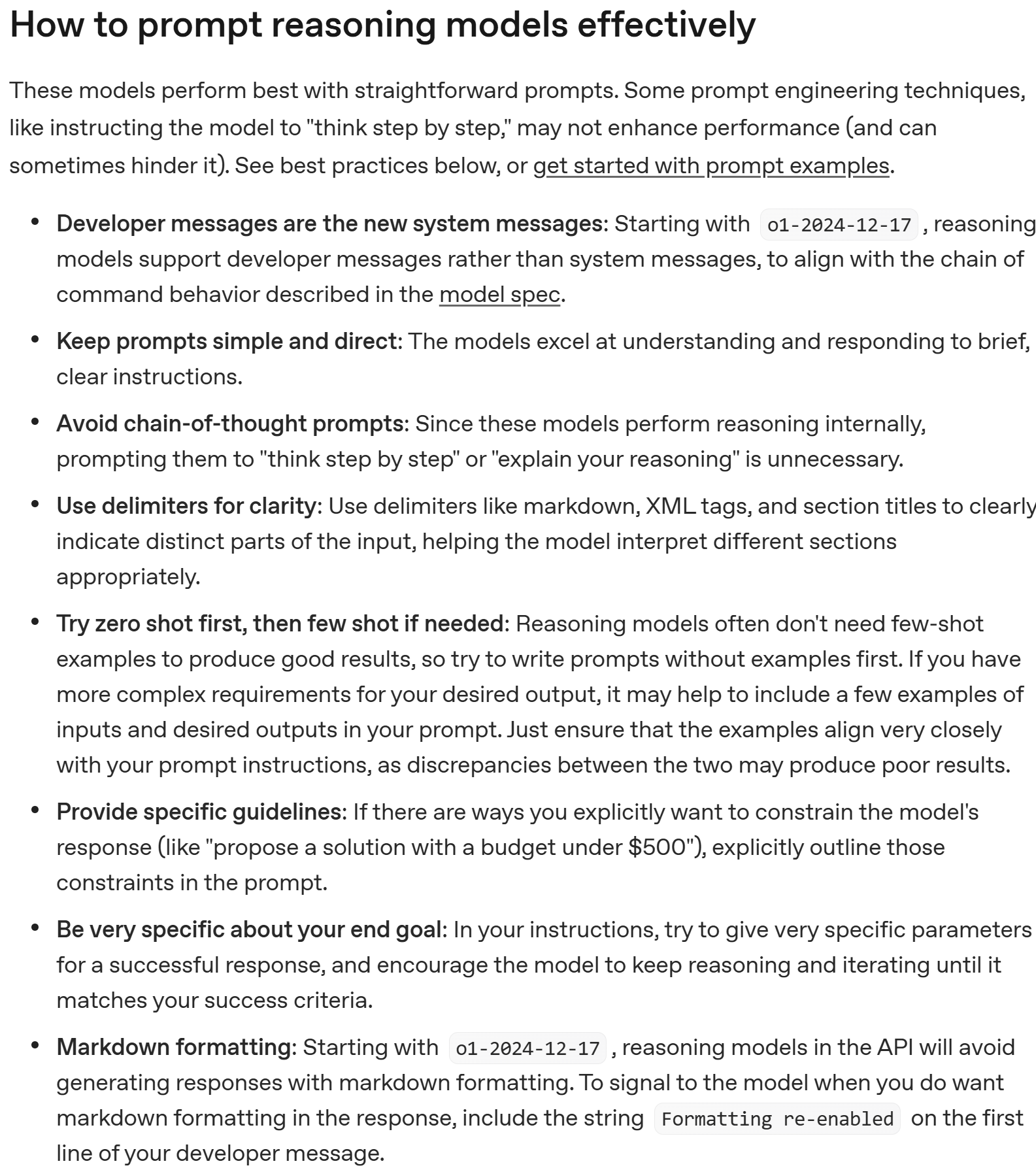}}
    \caption{Screenshot from Open AI's documentation webpage showing detailed prompt engineering advice for their reasoning models~\citep{OpenAIO1tWebpage}. Note how zero-shot chain-of-thought prompting is discouraged for these models.}
    \label{fig:openai-prompt-eng-advice}
\end{figure}

\textbf{Description.} An AI teammate should be able to understand intents at the same level that a senior software engineer would. Developers should not have to worry about phrasing their intents as prompts that must be crafted using an unnatural lingo that implements some specific \textit{prompt engineering}~\citep{PromptingGuide} technique (e.g., chain-of-thought~\citep{chainOfThought2022}). Moreover, research shows that prompts are very fragile~\citep{gao2021,hassan2024rethinking} (i.e., even slight variations in a prompt can lead to very different outputs). Finally, prompt engineering techniques do not generalize across models~\citep{chen2023chatgpts}, with different approaches proving effective for different architectures (Figure~\ref{fig:openai-prompt-eng-advice}).

\tinyskip \noindent \textbf{Affects.} The AI teammate as well as other layers of the entire stack (since every layer has FMware in it).

\begin{challengebox}{Open question \#6}
    \textbf{OQ6: }What is the best approach to avoid prompt engineering?

    \medskip \noindent \textbf{Our vision.} In our vision, the burden of crafting an effective prompt should be on the AI rather than on humans, be them users of IDE.next or the developers of the SE 3.0 stack. Although techniques such as Automatic Prompt Engineer~\citep{zhou2023large}, PromptBreeder~\citep{fernando2023promptbreeder} and DSPy~\citep{khattab2023dspy} exist, they still require complicated setup and are typically costly to use. A cost-effective solution like Compiler.next~\citep{CompilerNext25} is essential for seamlessly transforming human intent into optimized, model-specific prompts while maintaining efficiency and controlling costs. We foresee two promising research avenues:
 
    \tinyskip \noindent \textit{-- Novel model training strategies}. Frontier reasoning models such as OpenAI GPT 5.2 pro and DeepSeek V3.2 are trained with reinforcement learning techniques to perform complex reasoning. These models tend to require less prompt engineering by construction. In fact, certain prompt engineering directives such as ``think step by step'' tend to confuse those models and inadvertently degrade their performance. Promising research directions include reinforcement learning objectives that directly optimize for intent alignment under minimal instruction, curriculum learning strategies that expose models to progressively underspecified tasks, and reward signals that favor robustness to prompt paraphrasing.

    \tinyskip \noindent \textit{-- Building prompts from user feedback.} Another promising research avenue includes gathering human feedback (e.g., thumbs up/down) for prompt responses then automatically using that feedback to improve upcoming responses. For example, one could store \texttt{<instruction,response>} pairs where \texttt{response} is of good quality (e.g., it received a thumbs up) and later on leverage those pairs to create few-shot examples in the system prompt (note ``examples'' in Figure~\ref{fig:se3-dev}). Over time, with a big enough database, the model will learn to produce higher quality responses for the same prompt. Such a functionality is being currently developed by LangChain~\citep{langsmith_selflearn} and highlights the importance of collecting observability data. This solution can be further improved by combining it with \textit{crowdsourcing}, since many of the questions asked by developers are repetitive (fundamental principle behind Q\&A systems such as Stackoverflow). As developers interact with the AI Teammate to align their intents, we anticipate that they will eventually reach a wording that works well for their problems and give it a positive feedback. Such a \textit{battle-tested} prompt can also be stored in a vector database to create a \textit{template database}. Future prompts can be matched against that template and adjusted accordingly, a process also known as \textit{question calibration}. Companies such as Amazon have explored similar approaches~\citep{AmazonDevOpsGuru}.

\end{challengebox}

\subsection{Other open questions}

Last but not least, we list other open questions for which we have not yet developed a thorough vision yet: 

\begin{itemize}[wide = 0pt, topsep = 1.5pt, itemsep = 1.5pt]
    \item \textbf{OQ7:} What is a good software engineer in the SE 3.0 era? How should we train the next generation of software engineers? How should we reimagine the software engineering and computer science academic curricula for the SE 3.0 era? 
    
    \item \textbf{OQ8:} What will programming languages look like in the SE 3.0 era? How will human-readable languages for debugging, control, and legacy systems coexist and combine with new languages designed specifically for AI agents and FMs (e.g., token-efficient programming languages~\citep{Zhang2024})?

    \item \textbf{OQ9:} What should the UI of IDE.next look like? What would be the key elements of its interaction design? Should AI teammates or agents replace plugins? What does debugging look like in IDE.next? Do agents require their own specific IDEs~\citep{Hassan2025ASE}?
        
    \item \textbf{OQ10:} How can we evaluate the overall performance of Compiler.next~\citep{CompilerNext25} and what should the benchmark look like? How can we make such a benchmark actionable? More generally, how can we make FMs for code more interpretable~\citep{Denys2024}? 
    
    \item \textbf{OQ11:} Since AI Teammates are highly personalized and learn over time, what happens when a developer leaves a company? Is the teammate company's IP or a developer's personal asset? How to find a middleground? 
    
    \item \textbf{OQ12:} What is going to be the impact of SE 3.0 on job offerings? How will SE 3.0 impact the motivation of software engineers who are particularly fond of coding (both in general and specialized domains)?
    
    \item \textbf{OQ13:} How can we promote open innovation in the SE 3.0 era? How can we work as teams of researchers and innovators instead of siloed research labs?
    
    \item \textbf{OQ14:} Software should be \textit{for all and by all}~\citep{AIware}. How can we minimize the impact of SE 3.0 on accessibility, equity, and fairness due to its reliance on sophisticated infrastructure to develop, execute, and monitor FMware?
     
\end{itemize}

\section{Conclusion}
\label{sec:conclusion}


In SE 2.0, the software development process is task-driven, code-centric, and puts a significant cognitive overload on the human. As a consequence, despite the impressive capabilities of AI coding assistants, software development remains slow, error-prone, and costly. Bootstrapping developers' productivity while simultaneously creating high-quality software requires a deep rethinking of the role of AI in software engineering

In this paper, we discuss our vision of Software Engineering 3.0 (SE 3.0). SE 3.0 promotes (i) an intent-centric conversation-oriented development, (ii) an AI-native process, maximizing the strengths of human (e.g., to reflect about business requirements) and AI (e.g., ability to search for an optimal coding solution to a problem in hyper speed) in a symbiotic relationship, and (iii) the use of efficient knowledge-driven models with high reasoning capabilities. 

Given the breadth of our vision, one may wonder how the development of the five components of our proposed SE 3.0 technology stack (Figure~\ref{fig:se3-stack}) should be prioritized. As we discuss in Section~\ref{sec:challenges}, challenges exist in all those five components. Hence, to effectively advance the vision of SE 3.0, we suggest that those challenges should be researched in parallel, allowing progress in one area to inform and accelerate advancements in the others (e.g., similarly to the virtuous interplay between AI4SE and SE4AI). We acknowledge, however, that \textit{IDE.next} largely depends on all other components to be fully functional and useful. Ultimately, the SE 3.0 vision can only be truly assessed and validated as a whole once prototypes have been developed for all components of the stack.

Although we are already starting to see very early glimpses of SE 3.0 in the form of commercial vibe coding platforms (e.g., Lovable~\citep{Lovable}, Base44~\citep{Base44}, Replit~\citep{Replit}, Bolt.new~\citep{Bolt}, and V0 by Vercel~\citep{V0}), only time will tell whether our vision will come to life. Most importantly, we hope that this paper serves as a channel to (i) show that the advances in AI should make us rethink SE and (ii) trigger deeper discussions in the community (including opposing views), such that we can collectively reshape SE. In particular, we welcome official responses to this article.

We see the democratization of AI (OQ14) as a particularly important challenge that requires timely actions. The AI and SE communities should work together to create more cost-effective approaches for training FMs (e.g., inspired by \textit{FM.next}) and serving compound AI systems (e.g., inspired by \textit{Runtime.next}). We believe that designing ``small and mighty'' FMs with deeper SE knowledge and better reasoning capabilities is a particularly fruitful research area that is aligned with our vision and has the potential to drastically lower barriers to entry.

We finish this paper with a quote from the famous chess player Gary Kasparov many decades after his renowned game against IBM's Deep Blue AI~\citep{HBR2010}: ``A weak human player + machine + \textbf{better process} is superior to a very powerful machine alone, but more remarkably is superior to a stronger human player + machine + \textbf{inferior process}.'' The SE community must deeply rethink the SE process with AI at its core to ensure that AI can play a beneficial and transformative role in the future of software making.







\begin{acks}
    The findings and opinions expressed in this paper are those of the authors and do not necessarily represent or reflect those of Huawei and/or its subsidiaries and affiliates. Moreover, our results do not in any way reflect the quality of Huawei's products.
\end{acks}

\bibliographystyle{ACM-Reference-Format}
\bibliography{references}


\end{document}